\documentclass[prl,twocolumn,showpacs,aps,superscriptaddress]{revtex4-1}
 
\usepackage{amsmath}
\usepackage{amssymb}
\usepackage{amsmath}
\usepackage{txfonts}
\usepackage{graphicx}
\usepackage{epsfig}
\usepackage[T1]{fontenc}
\usepackage[latin9]{inputenc}
\usepackage{times}
\usepackage{color} 
\usepackage{xspace}
\usepackage{amssymb,amsmath}
\usepackage{amsbsy}
\usepackage{braket}
\usepackage{tabularx}
\usepackage{mathtools}          
\usepackage{ragged2e}
\usepackage[margin=1in]{geometry}
\usepackage{lipsum}
\usepackage{url}
\usepackage{enumitem}

\usepackage{ifpdf}
\ifpdf
\usepackage{epstopdf}
\fi
\usepackage[unicode]{hyperref}
\hypersetup{
	pdftitle={spinnorBEC},
	pdfauthor={Yu},
	pdfsubject={magneticdominawall},
	colorlinks=true,
	linkcolor=blue,
	citecolor=blue,
	filecolor=black,
	urlcolor=blue
}
\def\urlprefix{}
\def\url#1{}

\usepackage{bm}
\usepackage{color}

\newcommand{\be}{\begin{equation}}

\newcommand{\ee}{\end{equation}}

\newcommand{\bea}{\begin{eqnarray}}

\newcommand{\eea}{\end{eqnarray}}

\newcommand{\nn}{\nonumber}

\begin{document}

\justify


\title {Motion of ferrodark solitons in trapped superfluids: spin corrections and emergent oscillators}



\author{Jiangnan Biguo}
\affiliation{Graduate School of China Academy of Engineering Physics, Beijing 100193, China}
\author{Xiaoquan Yu}
\email{xqyu@gscaep.ac.cn}
\affiliation{Graduate School of China Academy of Engineering Physics, Beijing 100193, China}
\affiliation{Department of Physics, Centre for Quantum Science, and Dodd-Walls Centre for Photonic and Quantum Technologies, University of Otago, Dunedin 9016, New Zealand}

\begin{abstract}
We propose a framework for topological soliton dynamics in trapped spinor superfluids, decomposing the force acting on the soliton by the surrounding fluid into the buoyancy force and spin corrections arising from the density depletion at soliton core and the coupling between the orbital motion and the spin mixing, respectively.  
Our formulation applies to large-amplitude  soliton motion in general superfluids with spin degrees of freedom  under arbitrary external potentials.
For ferrodark solitons (FDSs) in  spin-1 Bose-Einstein condensates , the spin correction could diverge, change the direction of the total force and enable mapping the FDS motion in a harmonic trap to the atomic-mass particle dynamics in an emergent quartic potential. Initially placing a type-I FDS near the trap center,  a single-sided oscillation happens, which maps to the particle moving around a local minimum of the emergent double-well potential.  As the initial distance of a type-II FDS from the trap center increases, the motion exhibits three regimes: trap-centered harmonic and anharmonic oscallations followed by single-sided oscillations. Correspondingly the emergent  quartic potential undergoes a transition from a single minimum to a double-well shape, where the particle motion shifts from oscillating around the single  minimum to crossing between two minima via the local maximum,  then the symmetry-breaking motion around one of the two minima.  In a hard-wall trap with linear potential, the FDS motion maps to a harmonic oscillator.



 
\end{abstract}

\maketitle
\textit{Introduction--} Topological excitations associated with discrete symmetries, such as dark solitons, kinks, and domain walls,  exist in various systems and are of interest in many research areas~\cite{pitaevskii2016bose,chaikin1995principles, vachaspati2007kinks}. 
In superfluids, their dynamics reveals rich properties of superfluidity in the nonlinear regime.  In a  quasi-one-dimensional (1D)  harmonically trapped system where the system size is much larger than the soliton width, the ratio between the dark soliton oscillation frequency and the trap frequency is $1/\sqrt{2}$ for Bose gases~\cite{Busch2000,Pitaevskii2004,Collision_BEC2} and is approximately  $1/\sqrt{3}$ for unitary Fermi gases~\cite{Liao2011, Pitaevskii2011}.  Such results can be obtained from the  macroscopic equation of motion (EOM) $M_{\rm in} d^2X/d t^2=f_b$,  where $X$ is the soliton position, $f_b=\delta N d U/dX$ is the buoyancy force, $U(x)$ is the trap potential, $M_{\rm in}=2\partial \delta K/\partial V^2$ is the inertial mass, $\delta K$ is the soliton energy, $V=dX/dt$ is the soliton velocity,  and $\delta N$ is the atom number deficiency due to the density  depression at the core.   However,  the relation $\partial \delta K(\mu,V^2)/\partial \mu=\delta N$, an important step in the derivation from the energy conservation law $d\delta K/dt=0$,  was shown invalid for magnetic solitons in a two-component Bose-Einstein condensate (BEC) at finite $V$~\cite{Pitaevskii_2016_physical_mass},  indicating that this traditional formulation only applies to small amplitude oscillations~\cite{footnoteexception}. While a ferrodark soliton (FDS)~\cite{Yu2021,2022YX,2022CoreStructure,Yu2024snake} exhibits oscillations in a linear potential and the amplitude can be as large as the system size~\cite{2022YX}. Hence to describe large-amplitude soliton  oscillations involving spin degrees of freedom,  an alternative formulation is demanded.




In this Letter, we develop a general framework capable of describing the spin-dependent soliton motion  in a  quasi-1D trapped spinor superfluid and establish a mapping between the soliton motion and the dynamics of a particle with the atomic mass $M$ in an emergent potential.  Distinct from the traditional description~\cite{Pitaevskii2004, Pitaevskii2011},  in our formulation the soliton EOM reads $M_{\rm in} d^2 X/dt^2=f=f_b+f_s$, where the total force $f$ contains the buoyancy force $f_b$ and the additional spin correction $f_s$,  clarifying  the origin of the force acting on solitons in a  trapped system. The spin correction, which can be evaluated from the soliton energy explicitly, plays a key role in determining the spin-dependent soliton motion while being absent in scalar-soliton-like nonlinear waves.  The motion of  FDSs~\cite{2022YX} in a trapped spin-1 BEC provides the most appealing example of demonstrating the effects of the spin correction.  We find that $f_s$ and $f_b$ have opposite signs  for type-II FDSs and the total force $f$ could vanish and change sign  where $dU/dx\neq 0$, yields nontrivial equilibrium positions of the soliton motion.  At the transition point between type-I and type-II FDSs, both $f_s$ and $M_{\rm in}$ diverge and change sign while $f/M_{\rm in}$ keeps finite and nonzero.  During the motion, the buoyancy force $f_b$ remains  finite.  In a harmonic trap and a linear potential, the FDS motion maps to a  quartic oscillator exhibiting symmetry breaking and a harmonic oscillator, respectively.  Using this mapping, the oscillation frequencies are obtained   analytically utilizing exact FDS solutions~\cite{2022YX}.

\textit{Systems--} The mean-field Hamiltonian density  of a quasi-1D spin-1 BEC is \cite{MagneticOrder1,MagneticOrder2,MagneticOrder3,RevSpinorBEC, MagneticOrder5}: 
\bea
H =\frac{\hbar^2 |\partial_x \psi|^2}{2M} + \frac{g_n}{2}n^2+\frac{g_s}{2} |\mathbf{F}|^2+ U(x) n + q\psi^\dagger S_z^2 \psi,
\eea
where $\psi = (\psi_{-1},\psi_0, \psi_{+1})^T$ with components $\psi_{m=-1,0,1}$ denoting the amplitudes of magnetic sublevels, $M$ is the atomic mass, $n=\psi^\dagger \psi=\sum_m|\psi_m|^2=\sum_m n_m$ is the total  number density, $n_{m}$ is the component density, the magnetization $\mathbf{F}=\psi^\dagger \mathbf{S} \psi$ serves the order parameter associated with the  $\textrm{SO}(3)$ symmetry, $\mathbf{S} = (S_x, S_y,S_z)$, $S_{j=x,y,z}$ are spin-1 matrices,   $U(x)$ is the spin-independent potential, and $g_n>0$ and $g_s$ are the density and spin-dependent interaction strength, respectively.  The external magnetic field is along the $z$ axis and $q$ represents the quadratic Zeeman energy.  The dynamics of $\psi$ at zero temperature is governed by the spin-1 Gross-Pitaevskii equation (GPE):  $\partial \psi/\partial t=\delta H/\delta (i \hbar \psi^{\dagger})$.
For the ferromagnetic coupling ($g_s<0$) and $0<\tilde{q}\equiv-q/(2g_sn_b)<1$,  the uniform ground state is the easy-plane phase  under the constraint $F_z=0$ and is characterized by the nonvanishing transverse magnetization $ F_{\perp}\equiv F_x + i F_y$~\cite{RevSpinorBEC, MagneticOrder5},  where $n_b$ denotes the uniform total number density.


\textit{Ferrodark solitons--}
In the easy-plane phase,  FDSs are $\mathbb{Z}_2$ topological defects in the spin order (magnetic kinks) with positive (type-I) and negative (type-II) inertial mass.  
At  $g_s = -g_n/2$, the exact solutions are available and the transverse magnetization reads~\citep{2022YX}
$F_{\perp}^{\rm{I},\rm{II}}(x,t) =\sqrt{n_b^2 - q^2/g_n^2}\tanh[(x-Vt)/\ell^{\rm{I},\rm{II}}]$, where the width $\ell^{\rm I,II} = \sqrt{2\hbar^2/M(g_n n_b -MV^2 \mp Q)}$,  the minus (plus) sign in front of $Q$ specifies type-I (II) FDSs and $Q = \sqrt{M^2V^4 + q^2 -2g_n n_b MV^2}$. When $Q=0$, the speed limit $V=C_{\text{FDS}}=\sqrt{g_n n_b/M}\sqrt{1-\sqrt{1-\left( q/g_n n_b \right)^2}}$ is reached and the transition between two types occurs~\citep{2022YX}.  The soliton energy $\delta K=\int dx \, H[\psi_s]+\mu \delta N- \int dx \, H[\psi_g]=K_s-K_g$, 
where $K_s= \int dx \, (H[\psi_s]-\mu n) $, $K_g= \int dx \,(H[\psi_g]-\mu n_b)$, $\psi_s$ is the soliton wavefunction, $\delta N=\int dx \, (n_b-n_s)$ is the depleted atom number, $\psi_g$ is the ground state wave function, $n_s=|\psi_s|^2$, and  $\mu = (g_n + g_s)n_b + q/2$ is the chemical potential.  For FDSs, the total density $n^{\rm{I}, \rm{II}}$ shows a dip (but does not vanish) at the core and the soliton energy is  
\bea
\delta K^{\rm I,II}(n_b,\mu,V^2)=\frac{4 \hbar^4}{3g_nM^2 (\ell^{\rm I,II})^3}-\left(\frac{g_n n_b+q}{2}-\mu\right) \delta N^{\rm I,II}, \nn\\  
\label{energy}
\eea
where $\delta N^{\rm I,II}=2\hbar^2/g_n M \ell^{\rm I,II}$~\cite{footenotechemical}.

\textit{Reformulation of soliton dynamics and spin corrections--}
To incorporate the effect of spin mixing, here we propose a refined theory of soliton motion  in a  trapped superfluid.   Compared to the original theory developed in Refs.~\cite{Pitaevskii2004, Pitaevskii2011},  where the energy of a soliton in homogeneous systems is parametrized by $\mu$ and $V^2$,  the key difference of our formulation is parametrizing the soliton energy $\delta K(n_b,\mu,V^2)$ using $n_b$, $\mu$ and $V^2$ with $\mu$ appearing only in the term $\mu \delta N$.   Then within the local density approximation, the energy of a soliton in trapped superfluids can be obtained by taking the following substitution into $\delta K(n_b,\mu,V^2)$ :
\bea
 \mu &&\rightarrow \mu[X(t)]=\mu_N-U[X(t)], \\
n_b &&\rightarrow n_{b}[X(t)], \quad V \rightarrow \frac{d X(t)}{dt}, 
\eea 
where $X(t)$ is the position of the soliton core, $n_b(x)$ is ground state density profile for the trapped system, which can be obtained within the Thomas-Fermi (TF) approximation~\cite{pitaevskii2016bose}, and $\mu_N$ is determined by the total atom number $N$~\cite{footnoteTF}.   



 
Apparently  $\delta K\{n_b[X(t)],\mu[X(t)],V(t)^2\}$ is a function of $X(t)$ and $V(t)$ and we introduce $\widehat{\delta K} [X(t),V^2(t)]\equiv\delta K\{n_b[X(t)],\mu[X(t)],V(t)^2\}$.
The energy  conservation law $d \widehat{\delta K}[X(t),V^2(t)]/dt=(\partial \widehat{\delta K}/\partial X) (dX/dt)+(\partial \widehat{\delta K}/\partial V^2) (d V^2/dt)=0$ gives rise to 
\bea
		M_{\rm in} \frac{d^2X}{dt^2}=f=f_b+f_s,
\label{solitonEOM}
\eea
where $M_{\rm in}\equiv 2 \partial \widehat{\delta K}/\partial V^2$ is the soliton  inertial mass, and the total  force acting on the soliton $f \equiv -\partial \widehat{\delta K}/\partial X$ is decomposed into two parts,
\bea
\hspace{-10pt}f_b=-\frac{\partial \delta K}{\partial \mu(X)} \frac{\partial \mu (X)}{\partial X} \quad \text{and} \quad f_s=-\frac{\partial \delta K}{\partial n_b(X)} \frac{\partial n_b(X) }{\partial X}.
\label{forces}
\eea
%
Here, $\partial \delta K/\partial \mu=\int dx \, [n_b(x)-n_s(x)]=\delta N$ is valid for any velocity and $f_b= \delta N\partial U(X)/\partial X$ is the buoyancy force.  We denote the additional force $f_s$ as the spin correction.  For FDSs: 
\bea
f^{\rm I,II}_s=-\frac{\partial \delta K^{\rm I,II}}{\partial n_b(X)}\bigg|_{n_b=n^{\rm TF}_b}\frac{\partial n_b(X) }{\partial X}=\mp \frac{\hbar^2 V^2}{Q \ell^{\rm I,II}}\frac{\partial n_b(X) }{\partial X}, 
\label{spincorrection}
\eea
which is evaluated at the TF density~\cite{SM} and here  the  minus  (plus) sign  corresponds to type-I (II) FDSs.
Then the EOM of an FDS in a trapped spin-1 BEC reads
$M^{\rm I,II}_{\rm in}d^2X/dt^2=f^{\rm I,II}=f_b^{\rm I,II}+f_s^{\rm I,II}$, where 
\bea
\label{inertialmassFDS}
\hspace{-5mm}M^{\rm I,II}_{\rm in}&&=2 \frac{\partial \widehat{\delta K}^{\rm I,II}}{\partial V^2}\bigg|_{n_b=n^{\rm TF}_b}=\pm \frac{4\hbar^4}{g_n M Q (\ell^{\rm I,II})^3},\\
\hspace{-5mm}f^{\rm I,II}_b&&=-\frac{\partial \delta K^{\rm I,II}}{\partial \mu }\bigg|_{n_b=n^{\rm TF}_b}\frac{\partial \mu (X)}{\partial X}=-\frac{2\hbar^2}{g_n M \ell^{\rm I,II}}\frac{\partial \mu (X)}{\partial X},
\label{Buoyancy}
\eea
and  the  plus (minus) sign in Eq.~\eqref{inertialmassFDS} corresponds to type-I (II) FDSs. 

We emphasize that our formulation [Eqs.~\eqref{solitonEOM} and \eqref{forces}] applies to relevant solitons in general superfluids. For dark solitons in scalar BECs and dark-dark-dark solitons in spin-1 BECs~\cite{SM}, $f_s$ vanishes. While for magnetic solitons in two-component BECs~\cite{QuChunLei2016}, our approach gives rise to the same results~\cite{SM}  as reported in Ref.~\cite{Pitaevskii_2016_physical_mass} which were obtained  from a different perspective.

\textit{Equilibrium positions and the maximum speed--}  
The equilibrium positions occur where $a=f/M_{\rm in}=0$.  For FDSs, the total force $f^{\rm I,II}=(2\hbar^2/g_n M \ell^{\rm I,II})(1\pm M V^2/Q) \partial U(X) /\partial X$.  When approaching the type transition point $Q \rightarrow 0$, both $M^{\rm I,II}_{\rm in}$ and  $f_s^{\rm I,II}$ diverge ($f_b$ keeps finite) while  the ratio $f^{\rm I,II}/M^{\rm I,II}_{\rm in}$ remains finite and nonzero [Eqs.~\eqref{spincorrection} and~\eqref{inertialmassFDS}]. Then the equilibrium positions occur where the total force vanishes (with sign change), not where $M_{\rm in}$ diverges ($Q=0$). The total force vanishes either at $\partial U/\partial X=0$ or where $1-M V^2/Q=0$ while $\partial U/\partial X\neq0$ (for type-II FDSs). 
As it follows, during the FDS motion,  the maximum speed location (same as the equilibrium position $X_{\text{eq}}$) does not coincide with the position $X_*$ where the local speed limit ($Q=0$) is reached.    
Note that this maximum speed value is still less than the local speed limit $C_{\rm FDS}[n_b(x_{\text{eq}})]$ ~\cite{footnotespeedlimit}.  This scenario is illustrated explicitly in the case of a hard-wall confined system subjected to a linear potential $U(x) = kx$ (Fig.~\ref{FDSforce}).
\begin{figure}[h]
	\centering
	\hspace{0cm} 
		\includegraphics[width=0.5\textwidth,height=9.8cm]{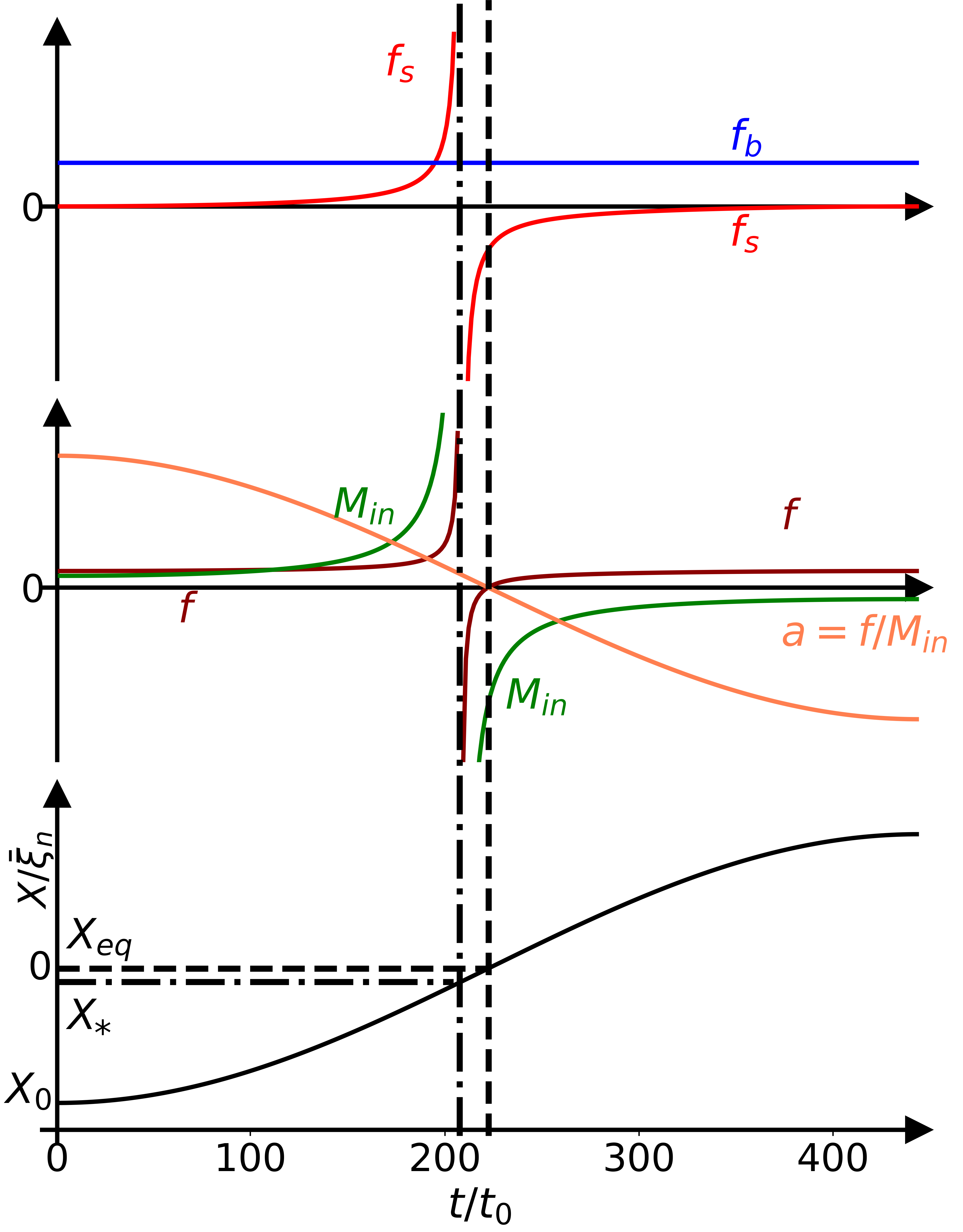}
	\caption{Evolution of the inertial mass $M_{\rm in}$ (green), the constant buoyancy force $f_b $ (blue), the spin correction $f_s$ (red),  the total force $f$ (dark red), and the FDS core trajectory (black solid)  during  a half period of the oscillation in a linear potential with  initially  placing a type-I FDS at $X_0$, evaluated using Eqs.~\eqref{spincorrection} ~\eqref{inertialmassFDS} and ~\eqref{Buoyancy}.  When crossing  the type transition point $X_{*}=\mu_N/k-\sqrt{(\mu_{N}/k-X_0-q/k)^2+q^2/4k^2}-q/2k$ (guided by the dash-dotted line), $M_{\rm in}$, $f_s$ and $f$ diverge and change sign.  The acceleration $a$ (orange) keeps sign before reaching the nearby equilibrium point $X_{\rm eq}=X_0+q/2k>X_{*}$(guided by the dash line), where the total force $f$ vanishes and changes sign again. Here  $X_0=-10\bar{\xi}_n$, $\tilde{q}=-q/2g_s\bar{n}_b=0.2$, $\mu_N=0.6 g_n \bar{n}_b$, $k=0.01 g_n \bar{n}_b/\bar{\xi}_n$ (for the parameters chosen, $X_{\rm eq}=0$),  $\bar{\xi}_n=\sqrt{\hbar^2/ M g_n\bar{n}_b}$, $t_0=\hbar/g_n \bar{n}_b$, and $\bar{n}_b$ is the average density.}
	\label{FDSforce}
\end{figure}

\textit{Emergent harmonic oscillators --} 
The FDS motion in a trapped spin-1 BEC  involves complex evolution of the inertial mass and the forces (Fig.~\ref{FDSforce}). On the other hand, remarkably, the EOM  of a FDS [Eq.~\eqref{solitonEOM}] maps to a dynamical system of a particle with the atomic mass: $dX/dt=\partial \tilde{H}/\partial P$, $dP/dt=-\partial \tilde{H}/\partial X$,
where $\tilde{H}=P^2/2M+\tilde{U}(X)$, $P=MdX/dt$ and the emergent potential 
\bea
\tilde{U}(X)=-\frac{2 \mu_N-4 E_0^{\rm I,II}-q}{ 4E^{\rm I,II}_0} U(X) +\frac{1}{4E^{\rm I,II}_0} U(X)^2.
\label{effectivepotential}
\eea
Here $E^{\rm I,II}_0=\hbar^2/2M (\ell^{\rm I,II})^2=\hbar^2/2M [\ell^{\rm I,II}(X_0,V_0)]^2$  is  determined by the initial position $X_0$ and the initial velocity $V_0$ of the soliton and serves as an energy unit.  The FDS motion in a linear potential described in Fig.~\ref{FDSforce} then maps to a harmonic oscillator and Eq.~\eqref{effectivepotential} becomes
\bea
\tilde{U}(X)=\frac{1}{2} M \omega_0^2 (X-X_{\rm eq})^2-\frac{1}{2} M \omega^2_0 X^2_{\rm eq},
\eea
where the equilibrium position $X_{\rm eq}=(2 \mu_N-4 E^{\rm I,II}_{0}-q)/2k=X_0\pm q/2k$ [plus (minus) sign specifies a type-I (II) FDS initially placed](coincides with the condition $f^{\rm II}=0$).  The  $X_0$-dependent oscillation frequency and the $X_0$-independent oscillation amplitude $A$ read~\cite{footenootenumerical}
\bea
\omega_0=\frac{|k|}{\sqrt{2 M E^{\rm I,II}_0}} \quad \text{and} \quad A=\frac{q}{2|k|}.
\eea

\begin{figure*}[htp]
	\centering
		\includegraphics[width=\textwidth,height=0.6\textwidth]{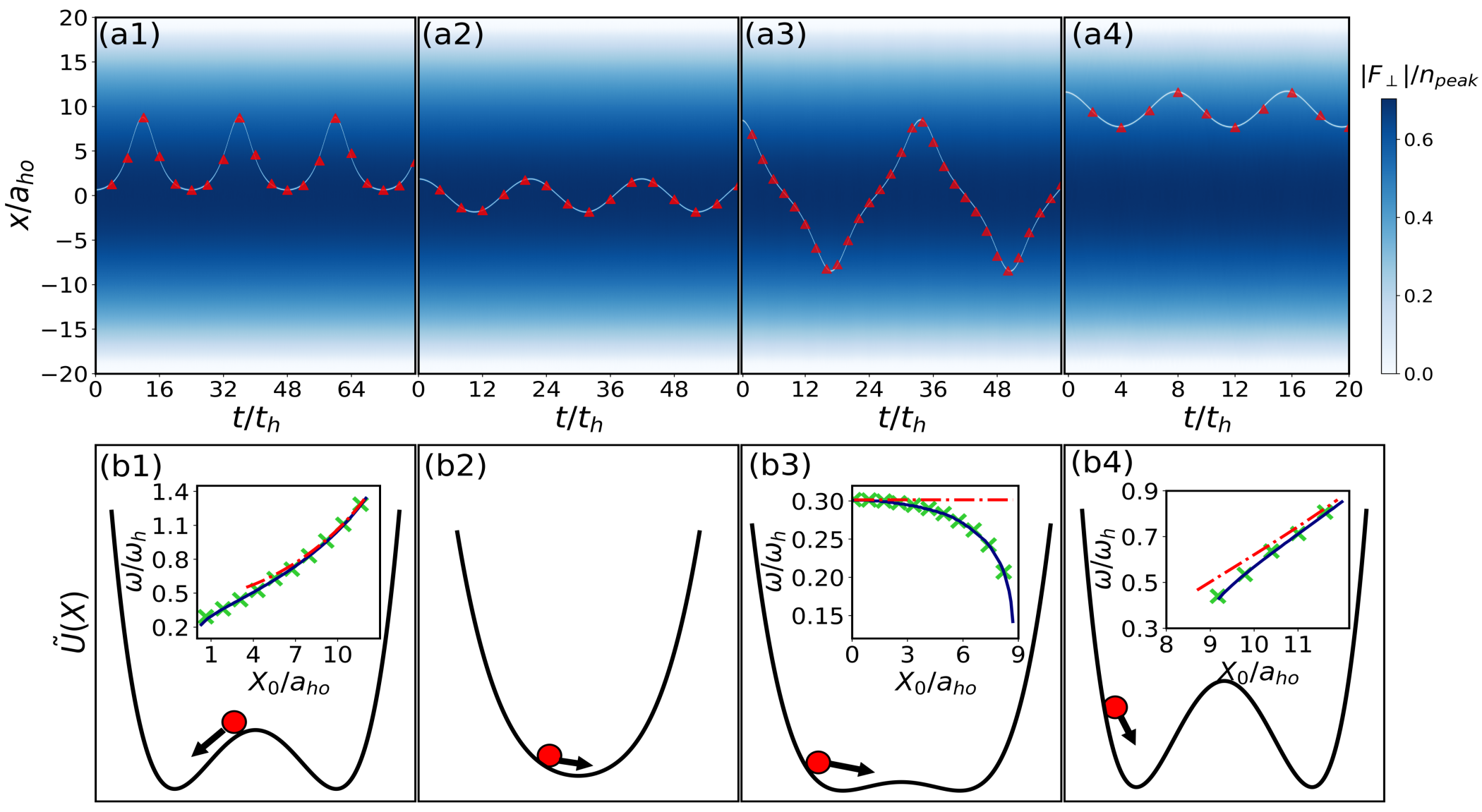}
	\caption{FDS dynamics in a harmonically trapped spin-1 BEC:  a FDS core trajectory  evolution[(a1)-(a4)] [spin-1 GPE simulations (lines); macroscopic soliton EOM Eq.~\eqref{solitonEOM} (markers)]  and the corresponding  emergent particle quartic potential $\tilde{U}(X)$ [(b1)-(b4)](schematic).  Here, $g_s/g_n =-1/2$, $\tilde{q} =-q/2g_s n_{\text{peak}}=0.1$, $\mu_N=201.2 \hbar \omega_h$, $a_{\rm ho} =\sqrt{\hbar/(M \omega_h)}$, $t_h = 1/\omega_h$ and $\omega_h$ is the trap frequency.  (a1) A type-I FDS is initially placed  near the trap center, it  drifts away from the trap center and then returns,  exhibiting  single-sided oscillations, mapping  to a particle rolling  down from the local maximum and then oscillating  around a local minimum of an emergent double-well potential [(b1)];  (a2)-(a4) When imprinting a type-II FDS, the motion depends on the initial distance to the trap center. (a2) Slightly away:  harmonic oscillations of a type-II FDS near the trap center,  mapping to particle oscillations around the single minimum of the emergent quartic potential [(b2)]; (a3) Further away:  anharmonic oscillations centered at the trap origin,  mapping to a particle going around the local maximum of the double-well potential [(b3)];  (a4) Far enough:   single-sided oscillations [inverse process of (a1)],  mapping to a particle moving  around a local minimum of the double-well potential [(b4)]. Insets in (b1)(b3) and (b4) show the initial position dependence of oscillation frequencies: analytical predictions for small  amplitude oscillations [Eqs.~\eqref{oscillationfrequencytypeI}, \eqref{oscillationfrequencyharmonic}   and \eqref{oscillationfrequencytypeII}](red dash-dotted lines) and large amplitude oscillations evaluated via $\omega = 2 \pi /\left|2\int^{X_{\rm end}}_{X_0} 1/V(X) \, dX\right|$ (black solid lines) with $X_{\text{end}}$ being the turning position, and numerical simulations of spin-1 GPEs (markers). }
	\label{oscillation_harmonic_fig}
\end{figure*}

\textit{Emergent quartic oscillators and symmetry-breaking trajectories--}  
For a harmonic potential $U(X)=M \omega^2 X^2/2$, 
\bea
\tilde{U}(X)=-\frac{1}{2} b M \omega^2 X^2+\frac{\lambda}{4} X^4,
\label{particlepotential}
\eea
where $b=(2\mu_N -4 E^{\rm I,II}_0-q)/ 4 E^{\rm I,II}_0$ and 
$\lambda=M^2  \omega^4/4  E^{\rm I,II}_0>0$. 
Hence the FDS motion  in a harmonically trapped spin-1 BEC  maps to a  quartic oscillator, a result of the combination of the buoyancy force and the spin correction\cite{footnotepotentialforscalar}.  

For a stationary  ($V_0=0$) type-I FDS  initially placed at $X_0$ near the origin, since $M^{\rm I}_{\rm in}>0$ and the force $f^{\rm I}$ points outwards, it starts to move toward the condensate boundary  and later undergoes a type transition,  leading to periodic motion on one side of the harmonic trap [Fig.~\ref{oscillation_harmonic_fig}(a1)] with the equilibrium position  located where the spin correction balances the buoyancy force ($1-M V^2/Q=0$).  Since $b=[2U(X_0)+q]/4 E^{\rm I}_0>0$,  $\tilde{U}(X)$ is a  double-well potential and the two minima locate at $X=X^{\rm I}_{\rm eq}=\pm \sqrt{X^2_0+q/M \omega^2}$ with $|X^{\rm I}_{\rm eq}|>X_0$~\cite{footnotestationary}.  This motion then maps to  a particle oscillating around one of the two local minima of the emergent potential $\tilde{U}(X)$ [Fig.~\ref{oscillation_harmonic_fig}(b1)].  To avoid reaching the phase boundary separating the easy-plane phase from the polar state,  the initial position of the FDS should satisfies  $|X_0|<\sqrt{(2 \mu_N-4q)/(M \omega^2)}$~\cite{footnoteinitialposition}.
When the oscillation amplitude is small, expanding $\tilde{U}$ around $X^{\rm I}_{\rm eq}$ we obtain $\tilde{U}=\tilde{U}''(X^{\rm I}_{\rm eq}) (X-X^{\rm I}_{\rm eq})^2/2$ and the oscillation frequency is  
\bea
\omega^{\rm I}=\sqrt{\frac{\tilde{U}''(X^{\rm I}_{\rm eq})}{M}}=\omega\sqrt{\frac{2[2U(X_0)+q]}{g_n n^{\rm TF}_b(X_0)-q}}.
\label{oscillationfrequencytypeI}
\eea

Starting with a stationary type-II FDS, three distinct trajectories are found:
(i) Initially placed at $X_0$ near the origin, the type-II FDS oscillates harmonically around the trap center [Fig.~\ref{oscillation_harmonic_fig}(a2)].  Since $b=[2 U(X_0)-q]/4 E^{\rm II}_0<0$ for sufficiently small $|X_0|$,  the emergent particle potential $\tilde{U}(X)$ exhibits a single minimum [Fig.~\ref{oscillation_harmonic_fig}(b2)].  For small amplitude oscillations ($|b M \omega^2| \gg \lambda X^2_0$), the oscillation frequency is obtained  by neglecting the quartic term~\cite{footnotepotentialphysicalmass}: 
	\bea
	\hspace{-1mm}\omega^{\rm II}_s = \sqrt{\frac{\tilde{U}''(0)}{M}}=\omega \sqrt{-b} =\omega \sqrt{\frac{q}{g_n n^{\rm TF}_b(0) +q}},
	\label{oscillationfrequencyharmonic}
	\eea
	which is independent of $X_0$. In this case the spin correction plays little role. 
As $|X_0| \rightarrow X^c_0$, where $X^c_0=\sqrt{q/M\omega^2}$ is determined by $b(X^c_0)=0$, $\omega^{\rm II}_s\rightarrow 0$, resembling the phenomenon of critical slowing down.  (ii) When $|X_0|>X^c_0$,  $b>0$ and $\tilde{U}(X)$ undergoes a  transition and becomes a   double-well potential with two minima  located at $X=X^{\rm II}_{\rm eq}=\pm\sqrt{X^2_0-q/M \omega^2}$ with $|X^{\rm II}_{\rm eq}|<|X_0|$ (balance between $f_b$ and $f^{\rm II}_s$: $1-M V^2/Q=0$).  When $X^c_0<|X_0|<\sqrt{2} X^c_0$, $\tilde{U}(X_0)>\tilde{U}(0)=0$, the mapped particle  motion is oscillatory  around the origin while  the trajectory crosses the maximum and the two minima of the potential $\tilde{U}(X)$ [Fig.~\ref{oscillation_harmonic_fig}(b3)].  This corresponds to the periodic motion of an FDS around the trap center while far from being  harmonic [Fig.~\ref{oscillation_harmonic_fig}(a3)]. (iii) For sufficiently large $|X_0|$, i.e., $|X_0|>\sqrt{2}X^c_0$~\cite{footnotephaseboundary}, $\tilde{U}(X_0)<\tilde{U}(0)=0$,  the symmetry-breaking oscillation around a local minimum $X=X^{\rm II}_{\rm eq}$ happens [Fig.~\ref{oscillation_harmonic_fig}(b4)], which corresponds to the periodic motion of an FDS on one side of the harmonic trap [Fig.~\ref{oscillation_harmonic_fig}(a4)].  Around $X^{\rm II}_{\rm eq}$, $\tilde{U}\simeq\tilde{U}''(X^{\rm II}_{\rm eq}) (X-X^{\rm II}_{\rm eq})^2/2$ and the oscillation frequency reads
	\bea
	\omega^{\rm II}=\sqrt{\frac{\tilde{U}''(X^{\rm II}_{\rm eq})}{M}}=\omega\sqrt{\frac{2[2U(X_0)-q]}{g_n n^{\rm TF}_b(X_0)+q}}.
	\label{oscillationfrequencytypeII}
	\eea
Note that as $q\rightarrow 0$, $X^c_0=0$, $X_{\rm eq}^{\rm I}=X_{\rm eq}^{\rm II}=X_0$, and $\omega^{\rm I}=\omega^{\rm II}\neq0$ (for $X_0 \neq 0$),  indicating the presence of  internal oscillations driven by the external force without orbital motion.  The relevant details will be discussed elsewhere. 

As discussed above,  by increasing $|X_0|$,  the emergent  particle potential $\tilde{U}$ undergoes a transition,  exhibiting a single minimum for $|X_0|<X^c_0$  and  two local minima  for $|X_0|>X^c_0$, where the critical value $X^c_0$ is determined by the flatness condition of the emergent quatic potential, i.e., $b=0$. This condition corresponds to the type-II FDS transferring to the type-I FDS right at the center of the harmonic trap. Indeed the onset condition for the type transition is $E^{\rm II}_{0}(X_0,0)=E^{\rm II}_{0}\{0,C_{\rm FDS}[n^{\rm TF}_b(0)]\}$, leading to
$|X_0|=\sqrt{2 \mu_N \left(1-\sqrt{1-q/\mu_N}\right)/M\omega^2}\simeq X^c_0$ for $q/\mu_{N} \ll 1$, which is fulfilled within the requirement of the TF approximation. 
Hence when $\tilde{U}$ becomes a double-well potential (ii and iii),  the FDS motion involves the type transition~\cite{2022YX}.

It is worthwhile mentioning that during oscillations the total number density profile of the FDS changes little  with respect to the local background density (see Fig.~S2~\cite{SM}).  While internal oscillations between $m=\pm 1$ and $m=0$ spin states take place at the core (Figs.~S3 and S4~\cite{SM}), driving transitions between type-I and type-II FDSs. Moreover,  we emphasize that away from the exactly solvable parameter region ($g_s=-g_n/2$, $0<q<-2g_s n_b$) the characteristic features of the FDS motion hold for other values of $g_s/g_n$ (Fig. S5~\cite{SM}).

\textit{Conclusion--} 
We formulate a theory of describing  spin-dependent soliton motion in a system confined by an external potential,  in which the total force  acting on the soliton is decomposed into the  buoyancy force and the spin correction. The interplay between the inertial mass, the buoyancy force and the spin correction leads to single-sided and trap-centered oscillations of ferrodark solitons in a harmonically trapped spin-1 BEC,  mapping to the motion of an atomic-mass particle in an emergent quartic potential.  Relevant frequencies and amplitudes of ferrodark soliton oscillations involving type transitions are obtained analytically within this mapping, which are otherwise challenging to calculate.   Our formulation is applicable to various nonlinear waves in trapped multicomponent superfluids~\cite{QuChunLei2016, Jieliu,chai2020magnetic,Oscillate_harmonic_trap, chai2022magnetic,2006YouL,Oscillate_constant_force, LCZhao_2, Peng2016, Mateo2022}.  Experimental investigations of  rich dynamics of  ferrodark solitons  in trapped systems are also within the scope of current ultracold-gas experiments~\cite{Dalibard2015,Gauthier16,Semeghini2018,Higbie2005,Huh2020a,chai2020magnetic,Oscillate_harmonic_trap,Pr2022Condensation,rabec2025bloch}.



\textit{Acknowledgments--}
We thank Y. Bai, R. Han, P.~B.~Blakie and C. Ma for useful discussions. X.Y. acknowledges support from the National Natural Science Foundation of China (Grant No. 12175215, Grant No. 12475041), the National Key Research and Development Program of China (Grant No. 2022YFA 1405300) and  NSAF (Grant No. U2330401).

\textit{Data availability--} 
The data that support the fundings of this article are openly available \cite{data_source}; embargo periods may apply.

\pagebreak
\widetext
\begin{center}
	\textbf{\large Supplemental Material for ``Motion of ferrodark solitons in trapped superfluids: spin corrections and emergent oscillators''}
\end{center}

\setcounter{equation}{0}
\setcounter{figure}{0}
\setcounter{table}{0}
\makeatletter
\renewcommand{\theequation}{S\arabic{equation}}
\renewcommand{\thefigure}{S\arabic{figure}}


%

	This Supplemental Material includes the derivation Thomas-Fermi approximation in spin-1 BECs, detailed calculations of spin corrections for scalar dark solitons,  dark-dark-dark solitons in spin-1 BECs and magnetic solitons in two-component BECs,  and other  necessary materials for supporting the main results presented in the main manuscript. 
	


	
	

\maketitle

\section{Thomas-Fermi approximation in spin-1 BECs}

In the easy-plane phase the ground state component densities are $n^b_{\pm 1}=(1-\tilde{q})n_b/4$ and $n^b_0=n_b(1+\tilde{q})/2$,  where  $n_b$ is the total number density.   In the presence of a harmonic trap $U(x) =M\omega^2 x^2/2$, when the harmonic length $a_{\rm ho} =\sqrt{\hbar/(M \omega)}$ is much larger than the spin healing length $\xi_s =  \hbar/\sqrt{M |g_s| n_b(0)}$, or equivalently  $\omega \hbar \ll |g_s|n_b(0)$, where $\omega$ is the trap frequency, the ground state wavefunction can be  obtained by neglecting the kinetic term in the Gross-Pitaevskii  equation \cite{pitaevskii2016bose, ground_length}, which were referred to as the  Thomas-Fermi approximation. 

\begin{figure}
	\center
	\includegraphics[width=10cm]{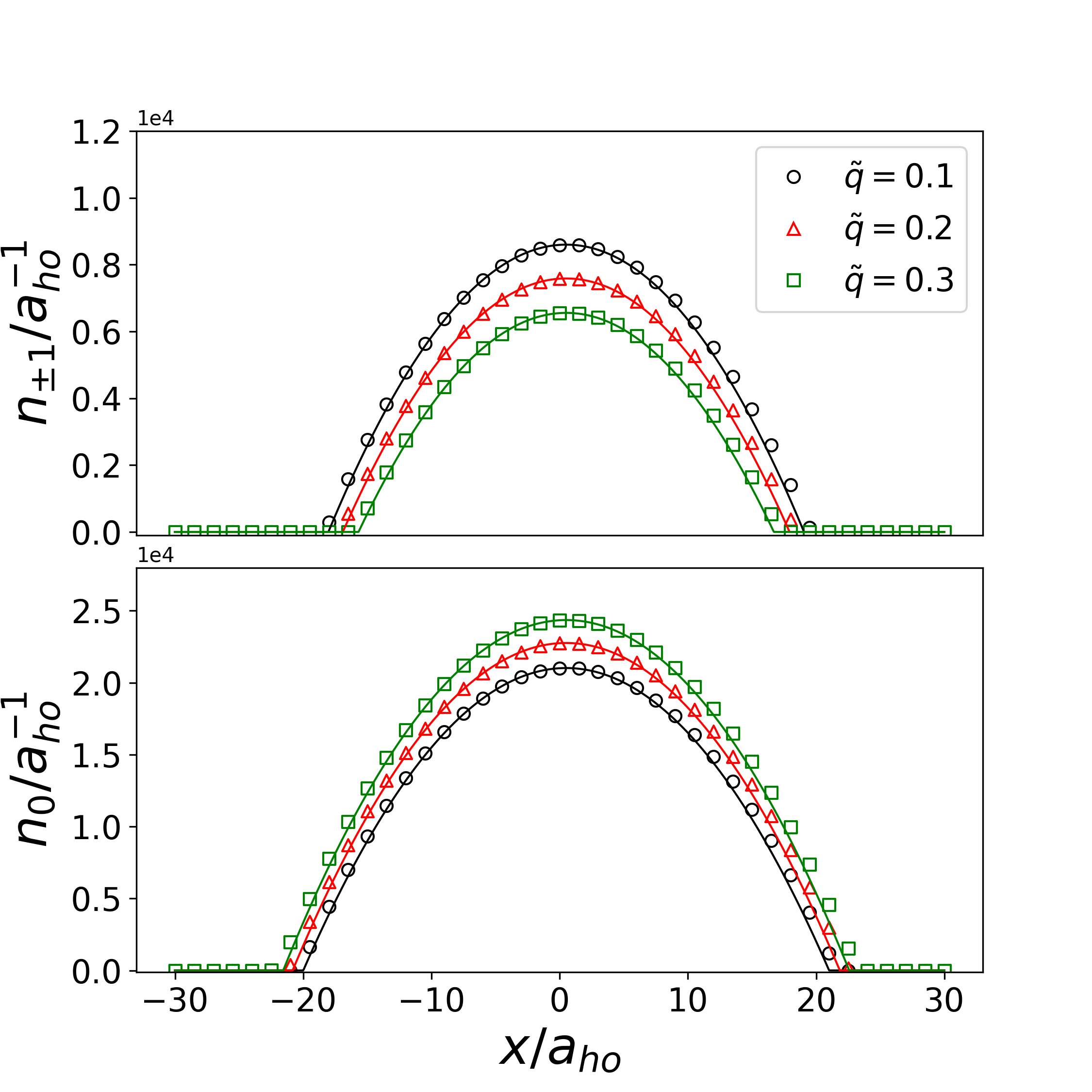}
	\caption{Comparison of three component densities  for a spin-1 BEC confined by  a harmonic trap $U(x)=M \omega^2 x^2/2$ between  the Thomas-Fermi approximation (solid lines) given by Eqs.~\eqref{TFdensity1} and~\eqref{TFdensity2} and numerical results obtained using the gradient flow  method~\cite{Bao2008a} (markers). Here $a_{\rm ho}= \sqrt{\hbar/M\omega}$,  $g_s/g_n = -1/2$,  the total particle number $N = 10^{6}$,  $\tilde{q} =-q/2 g_s n_{\text{peak}}= 0.1, 0.2,  0.3$,  and $n_{\text{peak}}$ is the total density at the trap center $x=0$.  }
	\label{TF}
\end{figure}

In the easy-plane phase,  within this Thomas-Fermi approximation, the component densities read
\bea
n^{\pm 1}_{\rm TF}(x)&=&\frac{\mu_{N}-U(x)}{4(g_n+g_s)} + \frac{g_n q}{8 g_s (g_n + g_s)}, \label{TFdensity1}\\
n^0_{\rm TF}(x)&=&\frac{\mu_{N}-U(x)}{2(g_n+g_s)} - \frac{(g_n + 2 g_s) q}{4g_s (g_n + g_s)}, \label{TFdensity2}
\eea
and  the total number  density reads 
\bea
n_{\rm TF}^{b}(x)&=&n^{+1}_{\rm TF}+n^{-1}_{\rm TF}+n^{0}_{\rm TF}=\frac{2\mu_{N}-2U(x)-q}{2(g_n+g_s)},
\label{TFdensity}
\eea
where the chemical potential $\mu_{N}$ is fixed by the normalization condition 
$\int dx \, n^b_{\rm TF}(x)=N$.  The condensate size $x_s$ is given by $n_{\rm TF}^{b}(x_s)=0$.  A comparison between the Thomas-Fermi densities [Eqs.~\eqref{TFdensity1} and \eqref{TFdensity2}] and numerical results which are obtained using the  gradient flow method~\cite{Bao2008a} is shown in Fig.~\ref{TF}. It is important to note that on the tails of the condensate, the density is low and the condition $0<q<-2g_s  n_{\rm TF}^b(x)$ may not be satisfied for given quadratic Zeeman energy $q$. For a harmonic trap, the phase boundaries $\pm x_b$ are determined by $q=-2g_s n_{\rm TF}^b(x_b)$ and  we have $x_b = \sqrt{(2\mu_N g_s+ g_n q)/(g_s M\omega^2)}$.
For $x_b<|x|<x_s$, the system is in the polar state~\cite{RevSpinorBEC,MagneticOrder5} and consistently we have $n_{\rm TF}^{\pm 1}(|x|\ge x_b)=0$. In the following we focus on the region $|x|<x_b$ where $n_{\rm TF}^{\pm 1}(x) >0$ and the system is in the easy-plane phase. 
For a  hard-wall trapped quasi-1D spin-1 BEC subjected to a linear potential $U(x) = kx$, the  TF density in the bulk  is 
\bea
n_{\rm TF}^b(x) =\frac{2\mu - 2kx -q}{2(g_n+g_s)}.
\eea
In the main manuscript  we only consider the situation where the Thomas-Fermi  condition is fulfilled. 

\vspace{5cm}

\section{Dynamics of the total density and component densities}

\begin{figure}[htp] 
	\centering
	\includegraphics[width=0.886\textwidth]{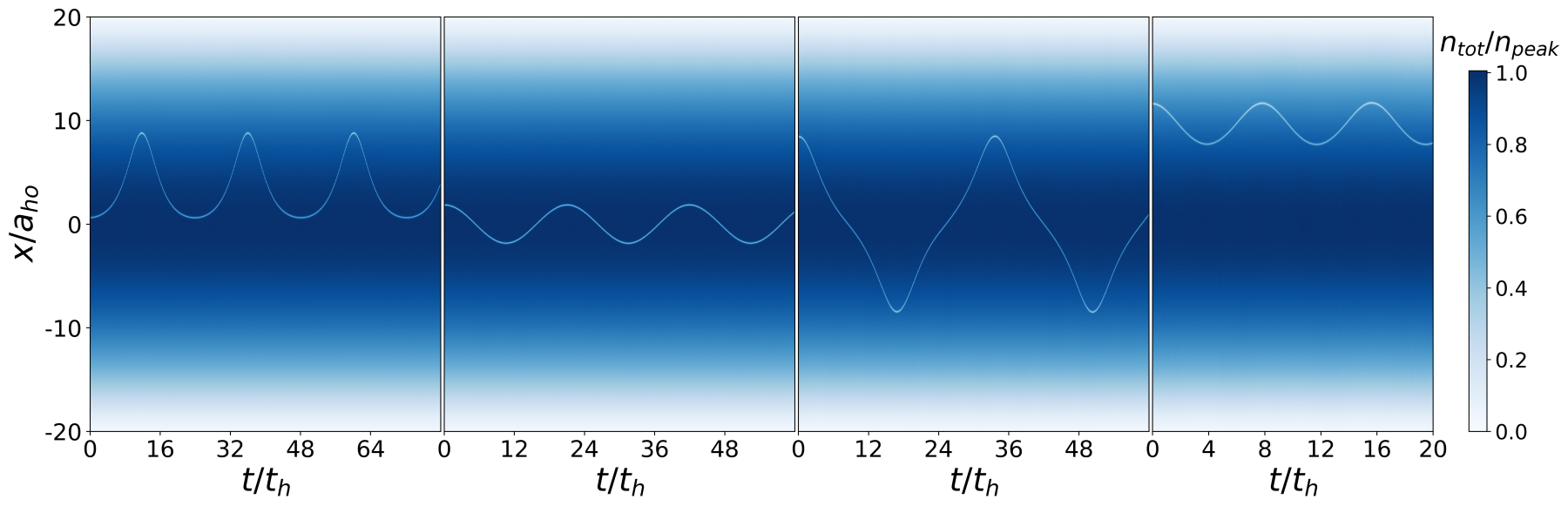}
	\caption{Time evolution of the total number density $n$. The  parameters are the same as those adopted in Fig.2.  
		\label{f:totaldensity}}
\end{figure}

\begin{figure}[htp] 
	\centering
	\includegraphics[width=0.886\textwidth]{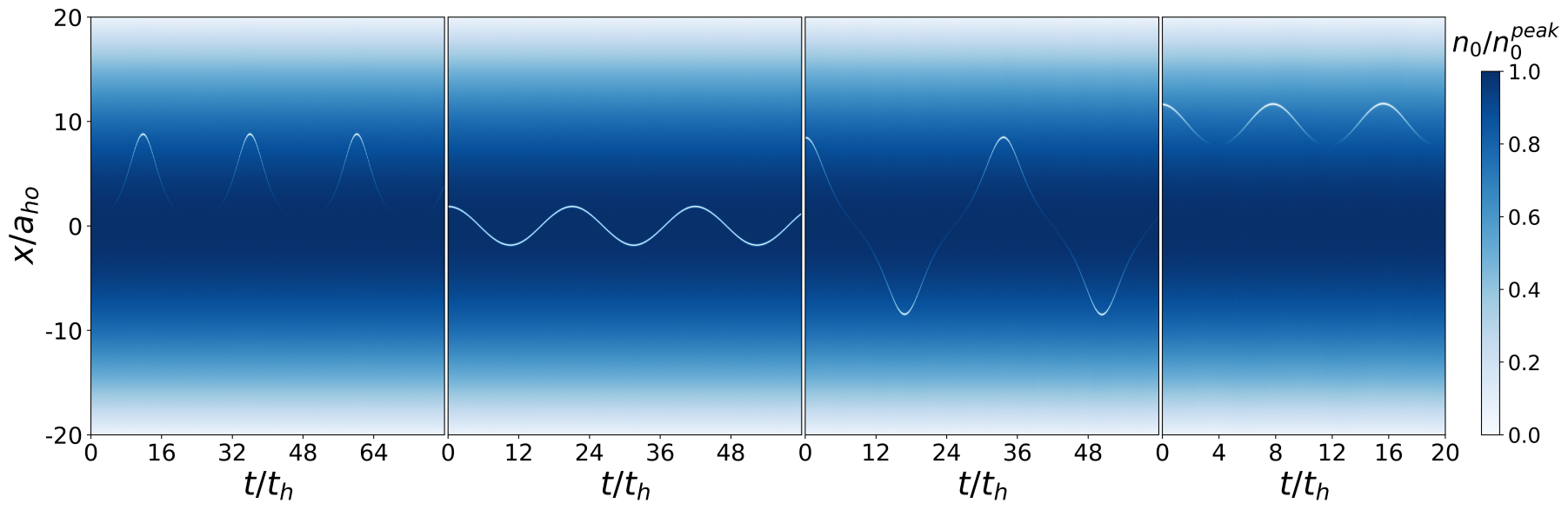}
	\caption{Time evolution of the component density $n_0$. The  parameters are the same as those adopted in Fig.2.  
		\label{f:componentdensityn0}}
\end{figure}

\begin{figure}[htp] 
	\centering
	\includegraphics[width=0.886\textwidth]{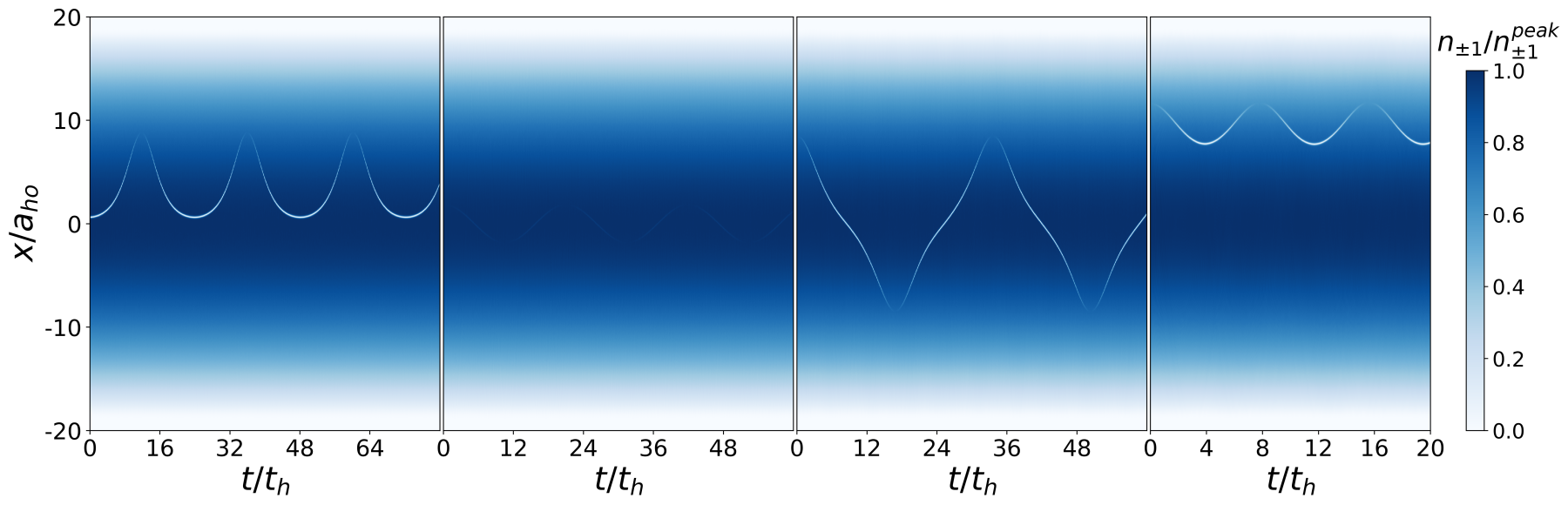}
	\caption{Time evolution of the component density $n_{\pm1}$. The  parameters are the same as those adopted in Fig.2. 
		\label{f:componentdensitynp}}
\end{figure}

During the FDS motion the total number density profile of the soliton has only minor changes with respect to the local background density (Fig.~\ref{f:totaldensity}).   However, internal oscillations between different  spin states near the core take place through the internal currents, inducing transitions between type-I and type-II FDSs.  Figures~\ref{f:componentdensityn0} and ~\ref{f:componentdensitynp} show the dynamics of $n_0$ and $n_p$, respectively.


\section{FDS motion away from the exactly solvable regime}
\begin{figure}[htp] 
	\centering
	\includegraphics[width=0.886\textwidth]{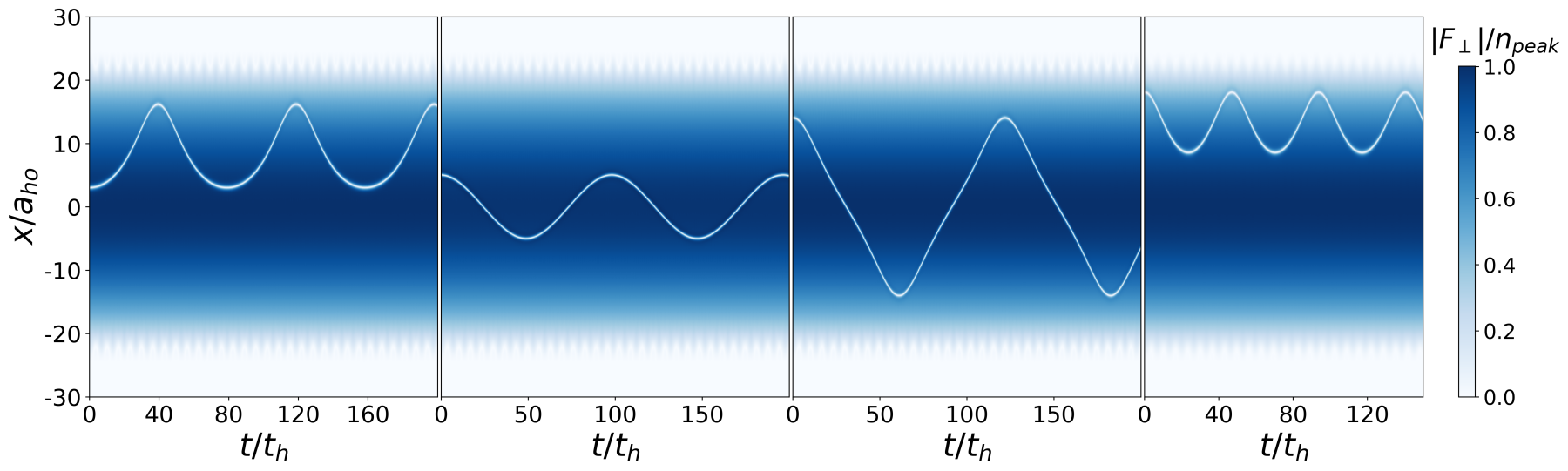}
	\caption{Dynamics of a ferrodark soliton in a harmonically trapped spin-1 BECs away from the exactly solvable regime.  The parameters are $g_s/g_n=-0.01$, $\tilde{q}=q/(-2g_s  n_{\rm peak})=0.2$, $\mu_N =299.69 \hbar \omega_{h}$, and the total particle number $N= 10^6$, where  $a_{\rm ho} =\sqrt{\hbar/(M \omega_h)}$, $t_h = 1/\omega_h$ and $\omega_h$ is the trap frequency.  The type-I ferrodark soliton is initially placed at $3a_{\rm ho}$. The type-II ferrodark soliton is initially placed at $5 a_{\rm ho}$,  $14 a_{\rm ho}$ and $18 a_{\rm ho}$, respectively.  
		\label{f:nonsolvble2}}
\end{figure}

The FDS dynamics revealed by the solution at the exactly solvable point holds in general.  Here we present the FDS oscillatory dynamics away from the exactly solvable regime and choose $g_s/g_n=-0.01$ which is very close to the value of $^{87}$ Rb.  As we can see that the dynamics is qualitatively the same as what in the exactly solvable regime (Fig.~\ref{f:nonsolvble2}). 

\section{Spin-corrections of scalar dark solitons and dark-dark-dark solitons}
In this section we show that for dark solitons in scalar BECs and dark-dark-dark solitons in spin-1 BECs the spin corrections vanish. 

The soliton energy  for a dark soliton in scalar BECs reads 
\bea
\hspace{-3mm}\delta K_{\rm scalar}=\frac{2 \hbar  \sqrt{\frac{g n_b(X)}{M}-V^2} \left[3 \mu(X) -g n_b(X)-2 M V^2\right]}{3 g},
\eea 
where $g$ is the  interaction strength. 

Following Eq. (6) in the main manuscript,  the spin correction for the scalar dark soliton is 
\bea
f_s=\frac{\partial \delta K_{\rm scalar}}{\partial n_b(X)}=\frac{\hbar  \left[\mu(X) -g n_b(X)\right]}{M \sqrt{\frac{g n(X)}{M}-V^2}}
\eea 
and it vanishes when taking the Thomas-Fermi density profile $n_b(x)=[\mu_{N}-U(x)]/g$.

Hence the equation of motion for a scalar dark soliton  is 
\bea
M_{\rm in} \frac{d^2X}{dt^2}=f_b,
\eea
where 
\bea
M_{\rm in}= 2 \frac{\partial \delta K_{\rm scalar }}{\partial V^2}\bigg|_{n_b=n^{\rm TF}}=-\frac{4\hbar M}{g} \sqrt{\frac{g n_b(X)}{M}-V^2}, 
\eea
and 
\bea
f_b=-\frac{\partial \delta K_{\rm scalar }}{\partial \mu(X)} \frac{\partial \mu (X)}{\partial X}\bigg|_{n_b=n^{\rm TF}}=\frac{2 \hbar }{g} \sqrt{\frac{g n_b(X)}{M}-V^2} \frac{\partial U (X)}{\partial X}.
\eea


In the easy-plane phase, there exist stationary topological defects in the density superfluid order associated with the  \textrm{U}(1) symmetry 
which are analogous to dark solitons in scalar BECs~\cite{pitaevskii2016bose} and are referred to as  dark-dark-dark solitons (DDDs)~\cite{Katsimiga_2021,2022CoreStructure}.  For $q=0$, we obtain exact solutions for finite velocity $V$:  
\bea
\psi_{0,\pm 1}=\sqrt{n_b^{0,\pm1}}\left[\sqrt{\frac{1-V^2}{c^2}} \tanh\left[\frac{(x-Vt)}{\xi}\right]+ i \frac{V}{c}\right],
\eea 
where the width $\xi=\hbar/M\sqrt{c^2 - V^2}$, $c = \sqrt{(g_n+g_s)n_b/M}$ is the group velocity of a gap-less low-lying mode at long wave-lengths~\cite{RevSpinorBEC,MagneticOrder5,2022YX}.  For this DDD soliton,  the transverse magnetization $F_{\perp}$ coincides with the total number density $n$ (or mass superfluid density) and  
\bea
F_{\perp}=n=n_b \left[ \left(1-\frac{V^2}{c^2}\right)\tanh^2\frac{ x-Vt}{\xi} + \frac{V^2}{c^2}\right]. 
\eea 
The excitation energy of such a soliton reads 
\bea
\delta K^{\rm D}[n_b, \mu, V^2] =\frac{4 \hbar^4} {3(g_n+g_s)M^2\xi^3} -[(g_n+g_s)n_b-\mu]\delta N,
\eea
where 
\bea
\delta N=\frac{2\hbar^2}{M (g_n+g_s) \xi}=\frac{2\hbar \sqrt{\frac{(g_n+g_s)n_b}{M}-V^2}}{g_n+g_s}.
\eea
Again we find that the spin correction for the DDD soliton    
\bea
\frac{\partial \delta K^{\rm D}}{\partial n_b(X)}=[\mu (X)-(g_n+g_s)n_b(X)] \xi
\eea
vanishes for the Thomas-Fermi density profile Eq.~\eqref{TFdensity}. 
Hence the equation of motion is 
\bea
M_{\rm in} \frac{d^2X}{dt^2}=f_b,
\eea
where 
\bea
M_{\rm in}=2 \frac{\partial \delta K^{\rm D}}{\partial V^2}\bigg|_{n_b=n^{\rm TF}}=-4 \hbar M\frac{\sqrt{\frac{n_b(X) (g_n+g_s)}{M} - V^2}}{g_n+g_s},
\eea
and 
\bea
f_b=-\frac{\partial \delta K_{\rm scalar }}{\partial \mu(X)} \frac{\partial \mu (X)}{\partial X}\bigg|_{n_b=n^{\rm TF}}=\frac{2\hbar \sqrt{\frac{(g_n+g_s)n_b(X)}{M}-V^2}}{g_n+g_s}  \frac{\partial U (X)}{\partial X}.
\eea

Then the  equations of motion for both scalar dark solitons and  DDD solitons  are  identical: 
\bea
M\frac{d^2 X}{dt^2} =-\frac{d \tilde{U}(X)}{dX}= -\frac{1}{2}\omega^2 M X,  
\label{DDD_EOM}
\eea
which maps to a particle dynamics in an emergent potential  $\tilde{U}(X)=(\omega/\sqrt{2})^2 M X^2/2$.   

%
%
%
%

%
%

\section{Spin corrections for magnetic solitons}
There exists so-called magnetic solitons in a two-component BEC~\cite{QuChunLei2016,Pitaevskii_2016_physical_mass} and the density difference $n_1-n_2$ has the meaning of spin polarization.  Within the formulation presented in the main manuscript,  the soliton energy is  
\bea
\delta K(n,\mu,V^2)=n \hbar \sqrt{\alpha\frac{\bar{g} n}{M}-V^2}+(g n-\mu) \delta N
\eea
and the depleted atom number
\bea
\delta N=\frac{3 \hbar}{2\bar{g}} \sqrt{c^2_s-V^2}. 
\eea
The buoyancy force 
\bea
f_b=-\frac{\partial \delta K}{\partial \mu(X)} \frac{\partial \mu (X)}{\partial X}=\frac{3\hbar}{2\bar{g}} \sqrt{c^2_s-V^2}\frac{\partial U(X)}{\partial X},
\eea
and the spin correction 
\bea
f_s=-\frac{\partial \delta K}{\partial n(X)} \frac{\partial n(X) }{\partial X}=\frac{\hbar}{2\bar{g}}\frac{V^2}{\sqrt{c^2_s-V^2}}\frac{\partial U(X) }{\partial X},
\eea
where $c_s=\sqrt{\alpha \bar{g} n/M}$ is the speed of ``spin sound '', $\alpha=\delta g/2\bar{g} \ll 1$, $\bar{g}=\sqrt{g_{11}g_{22}}$ and $\delta g=\bar{g}-g_{12}$. Here $g_{11}=g_{22}$ and $g_{12}=g_{21}$   refer to intra- and interspin coupling strength, respectively. 

Then the total force reads
\bea
f=\frac{\hbar}{\bar{g}}\left(\sqrt{c^2_s-V^2}+\frac{c^2_s}{2\sqrt{c^2_s-V^2}}\right) \frac{\partial U(X)}{\partial X}, 
\eea 
which is consistent with Eq.~(34) in Ref.~\cite{Pitaevskii_2016_physical_mass}.

\section{Oscillation frequency and amplitude of a FDS in a  linear potential}
\begin{figure}[h]
	\centering
	\includegraphics[width=9cm]{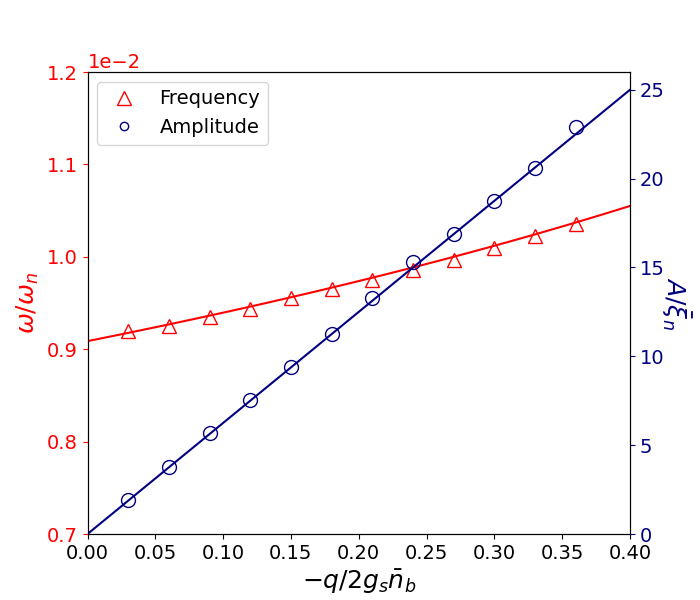}
	\caption{Comparison between analytical predictions (solid lines) and numerical results (markers) on oscillation frequency and amplitude of a FDS in a linear potential.  Here $g_s/g_n=-1/2$, the system size is $200 \bar{\xi}_n$,  $X_0=35\bar{\xi}_n$, $k=0.008 g_n \bar{n}_b/\bar{\xi}_n$, $\mu_N=g_n \bar{n}_b(1+\tilde{q})/2$, $\tilde{q}=-q/2g_s\bar{n}_b$, $\bar{\xi_n}=\hbar/\sqrt{Mg_n \bar{n}_b}$,  $\omega_n= g_n \bar{n}_b/\hbar$ and $\bar{n}_b$ is the average number density. }
	\label{fa}
\end{figure}

In the main manuscript,  we present the theoretical predictions of  the oscillation frequency and the amplitude  for a FDS in a hard-wall trapped system superimposed by a linear potential.  Here we show the analytical predictions against the spin-1 GPE  simulations of their dependence on the quadratic Zeeman  energy $q$ (see Fig.~\ref{fa}). 

%



\begin{thebibliography}{47}
\bibitem [{foo({\natexlab{j}})}]{footnotephaseboundary}%
	\BibitemOpen
	\href@noop {} {\emph {\bibinfo {title} {\rm{Note that $|X_0|<x_b$ has to be
					satisfied, where $x_b$ is the phase boundary seperating the easy-plane phase
					from the polar phase~\cite{SM}.}}}}\BibitemShut {Stop}%
\end{thebibliography}


\end{document}